\documentclass[twoside,fleqn]{article}
\usepackage{espcrc2}
\usepackage{graphicx}

\newcommand{\be}{\begin{equation}}
\newcommand{\ee}{\end{equation}}
\newcommand{\bea}{\begin{eqnarray}}
\newcommand{\eea}{\end{eqnarray}}

\newcommand{\cO}{{\cal O}}
\newcommand{\AmS}{{\protect\the\textfont2
  A\kern-.1667em\lower.5ex\hbox{M}\kern-.125emS}}

\title{ Vector Meson Dominance  as the first order \\
of a sequence of Pad\'e Approximants}

\author{J.J.~Sanz-Cillero~\address{
        Grup de F{\'\i}sica Te{\`o}rica and IFAE,
        Universitat Aut{\`o}noma de Barcelona, 08193 Barcelona, Spain.}
        \thanks{Talk given at the 14th International QCD Conference, QCD 08,
        7-12 July 2008, Montpellier (France). This work is based
        on Ref.~\cite{PadePeris} and has been done
        in collaboration with P.~Masjuan and S.~Peris.
I would like to thank the organisers  for the nice environment
during the conference.
This work has been supported by
CICYT-FEDER-FPA2005-02211, SGR2005-00916, the Spanish Consolider-Ingenio 2010
Program CPAN (CSD2007-00042)
and by the EU Contract No. MRTN-CT-2006-035482, ``FLAVIAnet''.
}
        }

\begin{document}

\begin{abstract}
The use of Pad\'e Approximants for the analysis of the pion vector form-factor \
is discussed and justified in this talk. The method is \
tested first in a theoretical model \
and applied then on real experimental data. \
It is shown how the Pad\'e Approximants  \
provide a convenient and reliable framework to incorporate both \
low and high energy information in the euclidean region, leading to improved determinations \
of the low energy parameters such as, e.g.,
the quadratic radius $\langle r^2\rangle^\pi_V$.
\end{abstract}

\maketitle

The object of the study of Ref.~\cite{PadePeris} was the pion
vector form-factor (VFF), defined by
\vspace*{-0.15cm}
\begin{equation}\label{def}
    \langle \pi^{+}(p')| \, J_{_{\rm E.M.}}^\mu \,  | \pi^{+}(p)\rangle
    \, =\,  (p+p')^{\mu} \ F(Q^2)\ ,
\end{equation}
\vspace*{-0.15cm}
with the electromagnetic current $J_{_{\rm E.M.}}^\mu=  \frac{2}{3}  \overline{u}\gamma^{\mu}u-\frac{1}{3}  \overline{d}\gamma^{\mu} d-
\frac{1}{3} \overline{s}\gamma^{\mu} s$  and $Q^2\equiv-(p'-p)^2$,
such that $Q^2>0$ corresponds to space-like data.

Our knowledge on  the VFF  from the time-like experimental data
tells us that the spectral function is essentially dominated by the
vector resonance $\rho(770)$. The remaining effects are almost negligible.
On the other hand, it is a well known fact that vector meson dominance
(VMD)  provides a good description of the space-like data
up to relatively high energies.

The question raised in this work is not so much ``why this is so''
but rather ``how can we make use of this knowledge''.
Our aim is to construct a space-like description of the VFF
that can be systematically improved. The Pad\'e Approximants (PAs)
provide this suitable framework.

A Pad\'e Approximant
to a given function is the ratio  of two polynomials $R_N(z)$ and $Q_M(z)$
(with degree $N$ and $M$, respectively), constructed such that the Taylor expansion
around some point (the origin in our case) exactly coincides with that of $F(z)$ up to the highest possible order, i.e.
$F(z)-P^N_M(z) =\cO(z^{M+N+1})$.  This may not look a big improvement
with respect to the usual Taylor power expansion
as they use the same inputs, the low energy coefficients. However, the
polynomial is unable to handle singularities such as
the logarithmic branch cut of the VFF,
whereas they are partially mimicked by the Pad\'e~\footnote{Notice that
if there were no singularities the Taylor expansion would converge everywhere
in the complex plane and the Pad\'es would not provide any further improvement.}.
%%The spectral function is actually
%%better and better approximated as the PA is considered at higher and higher orders.
Thus, in many cases
the Pad\'e works far beyond the range of applicability of Taylor expansions,
allowing the incorporation of both low and high energy data.  From this perspective,
VMD, where $F(Q^2)=(1+Q^2/M^2_{_{\rm VMD}})^{-1}$,  is just
a Pad\'e $P_1^0$, the first term of a sequence of the form $P^L_1$.

Our analysis  is focused on euclidean quantities
in real Quantum Chromodynamics (QCD), i.e., with $N_C=3$.
We separate away here from other studies based on QCD
in the limit of large number of colours~\cite{NC}
where the large-$N_C$ amplitudes are approximated through rational
functions~\cite{PereNC,PerisMasjuan08,PI:02}. In some situations, the poles
and residues were real and could be  related
to hadronic masses and couplings.
Instead, in the present work the Pad\'es are just a useful mathematical tool
which does not know about particles, masses, widths...
Our aim is to describe the euclidean momentum range and doing that
the Pad\'es tend to mimic the shape of the physical
spectral function -strongly peaked around the rho mass but finite-
through a finite number of poles.

The traditional point of view with respect to VMD has been that
the rho meson dominates, having the remaining resonances small couplings.
As we said, the PAs do not care about the particle poles in the amplitude.
The Pad\'e poles rather seem to be more related to bumps in the spectral
function than to
hadronic poles (masses, widths, couplings...).  The danger of relating
the poles of the rational approximants with mesonic resonances is clear.

Obviously, unlike the space-like region, we do not
expect to reproduce the time-like data
since a PA contains only isolated poles and cannot recover
a continuous time-like cut.

A final important point is that the PAs do not care about the asymptotic
behaviour of the amplitude at very short distances.
Although they work for a much wider range of momenta than a polynomial, the Pad\'e
eventually breaks down beyond some energy and they are not intended
to reproduce the VFF at $Q^2\to\infty$.

The importance of these techniques is  that they are able to produce
alternative outcomes competitive with other approaches.
The Pad\'es are  constructed in a simple and systematic way.
They provide an efficient procedure which allows incorporating high-energy
space-like  information.
This makes it an interesting alternative tool for the analysis of euclidean amplitudes.

There are several types of sequences of PAs
that may be considered. In order to achieve a fast numerical convergence, the
choice of which one to use is largely determined by the structure of the function to be
approximated. In this regard, a glance at the time-like data of the pion form factor makes it obvious that
the form factor is clearly dominated by the rho meson contribution. The effect of higher resonance states,
although present, is much more suppressed. In these circumstances the natural choice
is a single-pole  Pad\'{e} sequence $P^{L}_{1}$~\cite{Baker},
i.e. the ratio of a polynomial of degree $L$ over a polynomial of degree
one. Nonetheless, one should not confuse the Pad\'e pole
with the rho mass.

In order to test the aforementioned single-pole dominance,  we have also considered the
sequence $P^{L}_{2}$, confirming the results found with the PAs $P^{L}_{1}$.
We have also considered the so-called
Pad\'e-Type approximants (PTs)~\cite{PerisMasjuan08,math}
(rational approximants whose poles are fixed beforehand; here they have been
taken to be the physical
masses).  Finally, we also considered an intermediate case, the
so-called Partial-Pad\'e approximants (PPs)~\cite{math},
in which some of the poles are fixed to the physical masses  and some are left free
and fitted. The different results were rather independent of the kind of rational
approximant  used, being all of them  consistent among themselves.

\vspace*{-0.15cm}
\section*{Test with a model}
\vspace*{-0.15cm}

In order to illustrate the usefulness of the PAs,
we will first use a phenomenological model as a theoretical laboratory
to check our method.

A VFF phase-shift with the right
threshold behaviour is considered~\cite{Guerrero,Cillero,Portoles}:
\vspace*{-0.15cm}
\begin{equation}\label{model2}
    \delta(t)=\arctan \left[\frac{\hat{M}_{\rho}
    \hat{\Gamma}_{\rho}(t)}{\hat{M}_{\rho}^2-t} \right]\ ,
\end{equation}
\vspace*{-0.15cm}
with the $t$-dependent width given by~\cite{Guerrero,Cillero}
\vspace*{-0.15cm}
\begin{equation}\label{width}
    \hat{\Gamma}_{\rho}(t)= \Gamma_{0}\ \left( \frac{t}{\hat{M}_{\rho}^2} \right)\ \frac{\sigma^3(t)}{\sigma^3(\hat{M}_{\rho}^2)}\ \theta\left( t- 4 \hat{m}_{\pi}^{2} \right)\ ,
\end{equation}
\vspace*{-0.15cm}
and $\sigma(t)=\sqrt{1-4 \hat{m}_{\pi}^{2}/t}$.
The input parameters  are chosen to be close to their physical values
$ \Gamma_{0} = 0.15$~GeV, $\hat{M}_{\rho}^2= 0.6$~GeV$^2$,
$4 \hat{m}_{\pi}^{2}= 0.1$~GeV$^2$.
The form-factor
is then recovered through a once-subtracted Omn\'es relation,
\vspace*{-0.15cm}
\begin{equation}\label{model}
F(Q^2)= \exp\left \{-\frac{Q^2}{\pi}
\int_{4 \hat{m}_{\pi}^{2}}^{\infty}\ dt\ \frac{\delta(t)}{t (t+Q^2)}\right\}\ .
\end{equation}
\vspace*{-0.15cm}

At low energies, the form-factor is given by the Taylor expansion,
\vspace*{-0.15cm}
\begin{equation}\label{expmodel}
    F(Q^2)%%\approx
    \, =\, 1 \, + \, \sum_{k=1}^\infty  a_k\, (- Q^2)^k  \,\, ,
\end{equation}
\vspace*{-0.15cm}
where the coefficients $a_k$ are known since they are determined by
$\Gamma_0$, $\hat{M}_\rho^2$ and $4 \hat{m}_\pi^2$. The condition $F(0)=1$
has been already incorporated.

In order to recreate the experimental
situation~\cite{Amendolia}-\cite{Dally},  an emulation of the
space-like experimental
data is generated in the range $0.01$~GeV$^2 \leq Q^2\leq 10$~GeV$^2$~\cite{PadePeris}.

The Pad\'e Approximants $P^{L}_{1}(Q^2)$ are then fitted
to these euclidean ``data'' points, providing a prediction for the low-energy coefficients $a_k$.
It is found that as L increases the sequence of PAs $P^L_1$ converges
to the exactly known results, although
in a hierarchical way, i.e. much faster for $a_1$ than for $a_2$, and this one much faster than $a_3$,
etc... The relative error achieved in determining the coefficients $a_k$
by the Pad\'e  $P^4_1$  was,
respectively,  $1.5\%$ and $10\%$   for $a_1$ and $a_2$.
%%These results improve when the resonance width is decreased.
%%
%%It is also possible to build Pad\'e-Type
%%Approximants~\cite{PerisMasjuan08}, with
%%the $P^L_1$ pole has been placed at $s_p=\hat{M}^2_\rho$,
%%finding a similar convergence pattern.
%%
%%
These results will be taken  as a rough estimate of the systematic uncertainties
when fitting the real experimental data with Pad\'es in next section,
and they will be added to the final error~\cite{PadePeris}.

\vspace*{-0.15cm}
\section*{Experimental data analysis}
\vspace*{-0.15cm}

All the available experimental data in the space-like region have
been employed~\cite{Amendolia}-\cite{Dally},
ranging in momentum from $Q^2=0.015$ up to 10~GeV$^2$.

As discussed in the introduction, the prominent role of the rho meson contribution motivates
that we start  with the $P^{L}_{1}$ Pad\'e sequence.

\begin{figure}[!t]
\center
%\vspace{9pt}
%%\framebox[55mm]{\rule[-21mm]{0mm}{43mm}}
\includegraphics[width=7cm,clip]{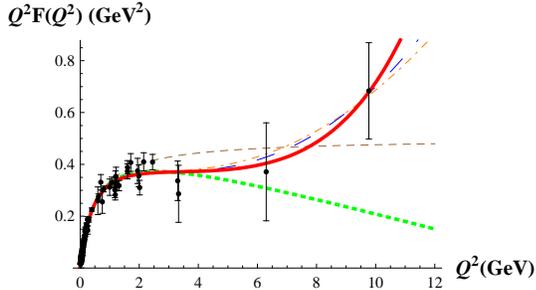}
\vspace*{-0.85cm}\caption{{\small The sequence of $P^L_1$ PAs is compared to the available space-like
  data~\cite{Amendolia}-\cite{Dally}:
  $P^0_1$ (brown dashed), $P^1_1$ (green thick-dashed),
  $P^2_1$ (orange dot-dashed), $P^3_1$ (blue long-dashed),
  $P^4_1$ (red solid).}}
\label{fig:VFF}
\end{figure}

The fit to $P^L_1$  yields a determination for the pion VFF and
the coefficients  $a_{k}$.
Nonetheless, according to Ref.~\cite{brodsky-lepage},
the VFF is supposed to  fall like $1/Q^2$ (up to logarithms) for $Q^2\to\infty$.
This means that, for any given value of $L$, one may expect   a good fit
only up to a finite value of $Q^2$, but not
for asymptotically large momentum.
This is clearly seen in Fig.~\ref{fig:VFF}, where the Pad\'{e} sequence
$P^L_1$ is compared to  the space-like  data. The Pad\'es converge to
the real VFF at low and mid energies but eventually diverge.

\begin{figure}[!t]
  \center
  \includegraphics[width=5.5cm]{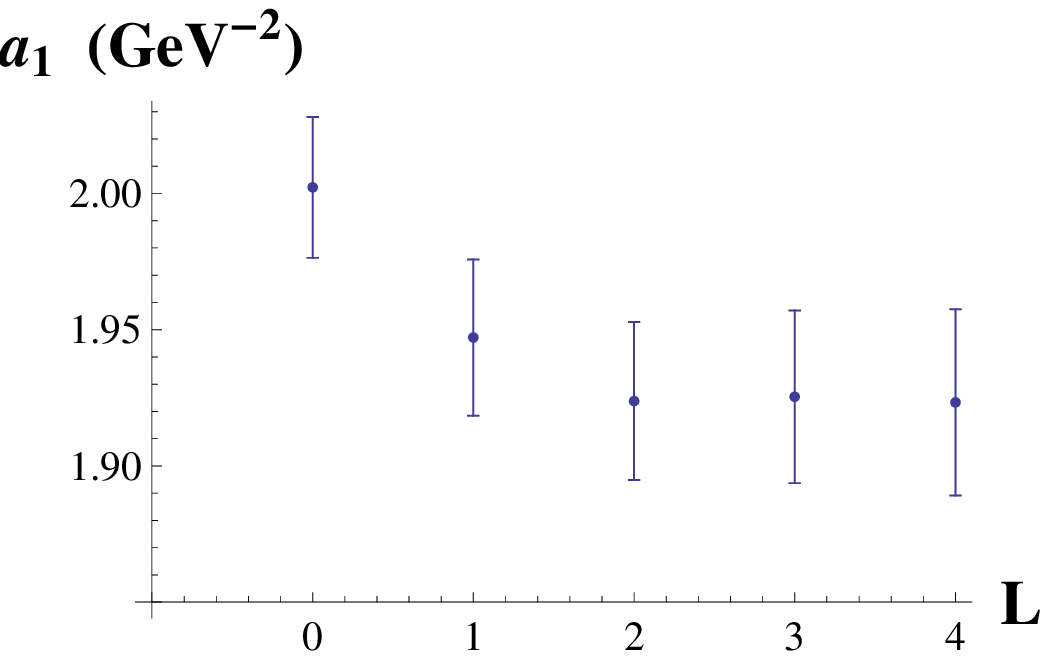} \\
%  \vspace*{0.3cm}
  \includegraphics[width=5.5cm]{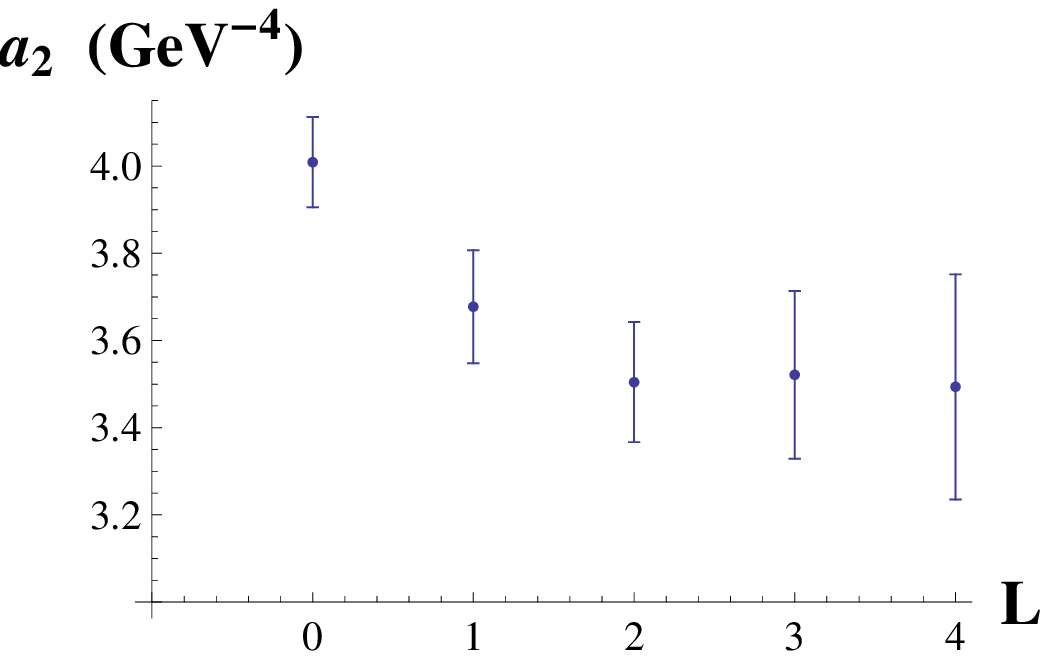}
\vspace*{-0.85cm}
  \caption{{\small $a_1$ and $a_2$ Taylor coefficients for the
  $P^L_1$ PA sequence.  }}\label{fig:a1PL1}
\end{figure}

Fig.~\ref{fig:a1PL1} shows the evolution of the fit results for the Taylor
coefficients $a_1$ and $a_2$.
As one can see, after a few Pad\'es they become stable~\cite{PadePeris}.
Thus, our best fit is provided by $P_1^4$~\cite{PadePeris}, yielding
\vspace*{-0.15cm}
\begin{equation}
\begin{array}{c}
a_1\, =\, 1.92 \pm 0.03\,\,\mbox{GeV}^{-2} \, ,\\
a_2\, =\, 3.49 \pm 0.26\,\,\mbox{GeV}^{-4} \,,
\end{array}
\end{equation}
\vspace*{-0.15cm}
with a $\chi^2/\mathrm{dof}=117/90$.

Alternative rational approximations were also constructed~\cite{PadePeris}:
two-pole PAs, Pad\'e-Types and Partial-Pad\'es~\cite{PerisMasjuan08}.
They all provided compatible results.

\vspace*{-0.15cm}
\section*{Conclusions}
\vspace*{-0.15cm}

\begin{table*}[!t]
\setlength{\tabcolsep}{1.5pc}
%%\newlength{\digitwidth} \settowidth{\digitwidth}{\rm 0}
\catcode`?=\active \def?{\kern\digitwidth}
%%%\begin{center}
\caption{{\small Our results for the quadratic radius $\langle r^2\rangle_V^\pi$ and second derivative $a_2$ are
compared to other
determinations~\cite{Portoles,Colangelo,ColangeloB,Yndurain,op6-VFF,lattice}.
Our first error is
statistical. The second one is systematic, based on the analysis
of the VFF model of the previous section~\cite{PadePeris}.}}
\label{results}
\begin{tabular*}{\textwidth}{@{}l@{\extracolsep{\fill}}ccc}
  \hline
   & $\langle r^2\rangle_V^\pi$   (fm$^2$)  &  $a_2$  (GeV$^{-4}$)     \\ \hline \hline
  This work~\cite{PadePeris} &   $0.445\pm 0.002_{\mathrm{stat}}\pm 0.007_{\mathrm{syst}}$ &  $3.30\pm 0.03_{\mathrm{stat}}\pm 0.33_{\mathrm{syst}}$    \\
  CGL~\cite{Colangelo,ColangeloB}& $ 0.435\pm 0.005$  & ...   \\
  TY~\cite{Yndurain}  & $ 0.432\pm 0.001 $  & $ 3.84\pm 0.02$ \\
  BCT~\cite{op6-VFF} & $0.437\pm 0.016$  & $3.85\pm 0.60$ \\
  PP~\cite{Portoles} & $0.430\pm 0.012$  & $3.79\pm 0.04$ \\
   Lattice~\cite{lattice} & $0.418\pm 0.031$  & ... \\
  \hline
\end{tabular*}
%%%\end{center}
\end{table*}

Combining all these outcomes, we obtained the results shown in Table~\ref{results}.
For comparison with previous analyses, the value
of the quadratic radius  is provided
(given by $\langle r^2 \rangle_V^\pi \, =\, 6 \, a_1$, with our best determination
$a_1=1.907\pm 0.010_{_{\rm sta}}\pm 0.03_{_{\rm sys}}$~GeV$^{-2}$).

In summary, in this work we have used rational approximations as a tool for
fitting the pion vector form factor in the euclidean range.
Since these approximants are capable of
going beyond the low-energy region, they are rather suitable for the
description of space-like data.
They allow to extract the low-energy coefficients, improving on their determination
by incorporating also the euclidean high-energy data information.
As one can see in Table~\ref{results},
the achieved degree of uncertainty is
shown to be competitive with previous analyses
existing in the literature~\cite{Portoles,Colangelo,ColangeloB,Yndurain,op6-VFF,lattice}.

%%\vspace{1cm}
%%
%%\textbf{Acknowledgements}

\end{document}